\documentstyle[11pt,paspconf,epsf]{article}

\def\gtrsim{\,\hbox{\hbox{$ > $}\kern -0.8em \lower
1.0ex\hbox{$\sim$}}\,}
\def\lesssim{\,\hbox{\hbox{$ < $}\kern -0.8em \lower
1.0ex\hbox{$\sim$}}\,}

\begin{document}

\title{Explosion Mechanism of Core-Collapse Supernovae ---\\
       a View Ten Years after SN~1987A}

\author{H.-Th.~Janka}
\affil{Max-Planck-Institut f\"ur Astrophysik, Karl-Schwarzschild-Str.~1,
       D-85740 Garching, Germany}

\begin{abstract}
The observation of neutrinos from Supernova~1987A has confirmed the 
theoretical conjecture that these particles play a crucial role during the 
collapse of the core of a massive star. Only one per cent of the
energy they carry away from the newly formed neutron star
may account for all the kinetic and electromagnetic energy
responsible for the spectacular display of the supernova explosion.
However, the neutrinos emitted from the collapsed stellar core at the center 
of the explosion couple so weakly to the surrounding matter that 
convective processes behind the supernova shock and/or inside the nascent
neutron star might be required to increase the efficiency of the energy
transfer to the stellar mantle and envelope. The conditions for a
successful explosion by the neutrino-heating mechanism and the possible
importance of convection in and around the neutron star are shortly reviewed.
\end{abstract}

\keywords{delayed explosion,convective instabilities,radiative transfer,
neutrinos}

\section{Introduction}\label{sec-1}

Even after ten exciting years SN~1987A keeps rapidly evolving and 
develops new, unexpected sides like an aging character. 
In the first few months the historical detection of
24 neutrinos in the underground facilities of the Kamiokande, IMB,
and Baksan laboratories caused hectic activity among scientists from very 
different fields. In the subsequent years the scene was dominated by the
rise and slow decay of light emission in all wavelengths which 
followed the outbreak of the supernova shock and contained a flood of
data about the structure of the progenitor star and the dynamics 
of the explosion. Now that the direct
emission has settled down to a rather low level, the supernova light
which is reflected from circumstellar structures provides insight into the 
progenitor's evolution. Even more information about the latter can be 
expected when the supernova shock hits the inner ring in a few years.

The neutrino detections in connection with SN~1987A were the
final proof that neutrinos take up the bulk of the energy during stellar 
core collapse and neutron star formation. Lightcurve and spectra
of SN~1987A bear clear evidence of large-scale mixing in the stellar
mantle and envelope and of fast moving Ni clumps (see, e.g., J.~Spyromilio,
D.~Wooden, K.~Nomoto, this volume). Both might indicate that macroscopic 
anisotropies and inhomogeneities were already present near the formation
region of Fe group elements during the very early stages of the explosion.
Spherically symmetric models had been suggesting for some time already
that regions inside the newly formed neutron star and in the neutrino-heated
layer around it might be convectively unstable. These theoretical
results and the observational findings in SN~1987A were motivation to 
study stellar core collapse and supernova explosions with multi-dimensional
simulations. 

In this article convective overturn in the neutrino-heated region
around the collapsed stellar core is discussed concerning its effects on 
the neutrino energy deposition and its potential importance for the 
supernova explosion. Convective activity inside the proto-neutron star is 
suggested as a possibly crucial boost of the neutrino luminosities on a
timescale of a few hundred milliseconds after core bounce. The first 
two-dimensional simulations that follow the evolution of the nascent 
neutron star for more than one second are shortly described.
\begin{figure}
\plotone{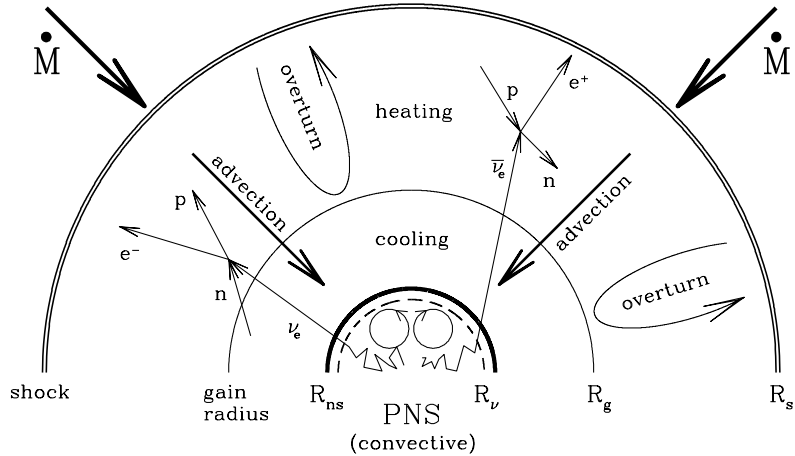}
\caption{Sketch of the post-collapse stellar core during the neutrino heating
and shock revival phase.}
\label{fig-1}
\end{figure}

\section{Neutrino-driven explosions and Convective overturn}\label{sec-2}
 
Convective instabilities in the layers adjacent to the nascent neutron star 
are a natural consequence of the negative entropy gradient built up by 
neutrino heating (Bethe 1990) and are seen in recent two- and three-dimensional
simulations (Burrows et al.~1995; Herant et al.~1992, 1994; 
Janka \& M\"uller 1995, 1996; Mezzacappa et al.~1997; Miller et al.~1993;
Shimizu et al.~1994). Although there is general
agreement about the existence of this unstable region
between the radius of maximum neutrino heating (which is very close outside
the ``gain radius'' $R_{\rm g}$, i.e.~the radius where neutrino cooling 
switches into net heating)
and the shock position $R_{\rm s}$, the strength of the convective overturn 
and its importance for the success of the neutrino-heating mechanism in
driving the explosion of the star is still a matter of vivid debate.

The effect of convective overturn in the neutrino-heated region on the shock
is two-fold. On the one hand, heated matter from the region close to
the gain radius rises outward and at the same time is replaced by cool 
gas flowing down from the postshock region. Since the production reactions
of neutrinos ($e^{\pm}$ capture on nucleons and thermal processes) are
very temperature sensitive, the expansion and cooling of rising plasma
reduces the energy loss by reemission of neutrinos. Moreover, the net energy
deposition by neutrinos is enhanced as more cool material is exposed to the 
large neutrino fluxes just outside the gain radius where the neutrino 
heating rate peaks (the radial dilution of the fluxes roughly goes as $1/r^2$).
On the other hand, hot matter floats into the postshock region and increases
the pressure there. Thus the shock is pushed further out which leads
to a growth of the gain region and therefore also of the net energy transfer 
from neutrinos to the stellar gas.

Figure~\ref{fig-1} displays a sketch of the neutrino cooling and heating 
regions outside the proto-neutron star at the center. The main processes
of neutrino energy deposition are the charged-current reactions 
$\nu_e+n \rightarrow p + e^-$ and $\bar\nu_e+p \rightarrow n + e^+$.
The heating rate per nucleon ($N$) is approximately
\begin{equation}
Q_{\nu}^+\,\approx\, 110\cdot {L_{\nu,52}\langle\epsilon_{\nu,15}^2\rangle
\over r_7^2 \,\, f}\cdot\left\{\matrix{Y_n\cr Y_p \cr}\right \}\quad
\left\lbrack {{\rm MeV}\over {\rm s}\cdot N}\right\rbrack \; ,
\label{eq-1}
\end{equation}
where $Y_n$ and $Y_p$ are the number fractions of free neutrons and protons, 
respectively, $L_{\nu,52}$ denotes the luminosity of $\nu_e$ or $\bar\nu_e$ in
$10^{52}\,{\rm erg/s}$, $r_7$ the radial position in $10^7\,{\rm cm}$,
and $\langle\epsilon_{\nu,15}^2\rangle$ the average of the squared neutrino 
energy measured in units of $15\,{\rm MeV}$. $f$ is the angular dilution factor
of the neutrino radiation field (the ``flux factor'', which is equal to the mean
value of the cosine of the angle of neutrino propagation relative to the
radial direction) which varies 
between about 0.25 at the neutrinosphere and 1 for radially streaming
neutrinos far out. Using this energy deposition rate, neglecting loss due to
re-emission of neutrinos, and assuming that the gravitational binding energy
of a nucleon in the neutron star potential is (roughly) balanced by the sum 
of internal and nuclear recombination energies after accretion of the infalling
matter through the shock, one can estimate the explosion energy to be of the
order
\begin{equation}
E_{\rm exp} \,\approx\, 2.2\cdot 10^{51}\cdot
{L_{\nu,52}\langle\epsilon_{\nu,15}^2\rangle\over r_7^2\,\, f}\,
\left({\Delta M \over 0.1\,M_\odot}\right)
\left({\Delta t\over 0.1\,{\rm s}}\right)
-\,E_{\rm gb} +\,E_{\rm nuc} \quad \left\lbrack \rm erg\right\rbrack \, .
\label{eq-2}
\end{equation}
$\Delta M$ is the heated mass, $\Delta t$ the typical heating timescale,
$E_{\rm gb}$ the (net) total gravitational binding energy of the overlying,
outward accelerated stellar layers, and $E_{\rm nuc}$ the additional energy
from explosive nucleosynthesis which is typically a few $10^{50}\,{\rm erg}$
and roughly compensates $E_{\rm gb}$ for progenitors with main sequence 
masses of less than about $20\,M_{\odot}$ (see also Burrows, this volume).
Since the gain radius, shock radius, and $\Delta t$ and thus also $\Delta M$
depend on $L_{\nu}\langle\epsilon_{\nu}^2\rangle$, the sensitivity of 
$E_{\rm exp}$ to the neutrino emission parameters is even stronger than 
suggested by Eq.~(\ref{eq-2}). 
\begin{figure}
\plotone{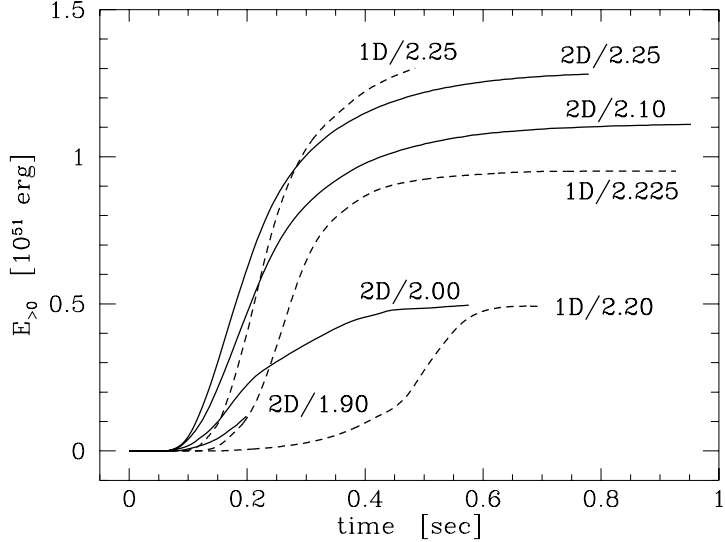}
\caption{Explosion energies $E_{>0}(t)$ 
for 1D (dashed) and 2D (solid) simulations with different assumed 
$\nu_e$ and $\bar\nu_e$ luminosities (labels give values in $10^{52}\,$erg/s)
from the proto-neutron star. Below the smallest given luminosities the
considered $15\,M_{\odot}$ star does not explode in 1D and acquires 
too low an expansion energy in 2D to unbind the stellar mantle and envelope.}
\label{fig-2} 
\end{figure}

In order to get explosions by the delayed neutrino-heating mechanism, certain
conditions need to be fulfilled. Expansion of the postshock region requires
sufficiently large pressure gradients near the radius $R_{\rm cut}$ of the
developing mass cut. If one neglects self-gravity of the gas in this region and
assumes the density profile to be a power law, $\rho(r) \propto r^{-n}$
(which is well justified according to numerical simulations
which yield a power law index of $n \approx 3$; see also Bethe 1993),
one gets $P(r)\propto r^{-n-1}$ for the pressure, and outward
acceleration is maintained as long as the following condition for the
``critical'' internal energy density $\varepsilon$ holds:
\begin{equation}
\left. {\varepsilon_{\rm c}\over GM\rho/r}\right|_{R_{\rm cut}} \,>\,
{1\over (n+1)(\gamma-1)}\,\cong\,{3\over 4} \; ,
\label{eq-3}
\end{equation}
where use was made of the relation $P = (\gamma-1)\varepsilon$.
The numerical value was obtained for $\gamma = 4/3$ and $n = 3$. This 
condition can be converted into a criterion for the entropy per baryon,
$s$. Using the thermodynamical relation for the entropy density
normalized to the baryon density $n_b$,
$s = (\varepsilon + P)/(n_b T) - \sum_i \eta_i Y_i$
where $\eta_i$ ($i = n,\,p,\,e^-,\,e^+$) are the particle
chemical potentials divided by the temperature, and assuming
completely disintegrated nuclei behind the shock so that the
number fractions of free protons and neutrons are $Y_p = Y_e$ and
$Y_n = 1 - Y_e$, respectively, one gets 
%
\begin{equation}
s_{\rm c}(R_{\rm cut}) \,\gtrsim\,
\left. 15\,\,{M_{1.1}\over r_7 \, T}\,\right|_{R_{\rm cut}}\,-\,
\left. \ln\left(1.27\cdot 10^{-3}\,\,{\rho_9\, Y_n\over T^{3/2}}\right)
\right|_{R_{\rm cut}}\quad \left\lbrack k_{\rm B}/N\right\rbrack 
\;.
\label{eq-4}
\end{equation}
In this approximate expression a term with a factor $Y_e$
was dropped (its absolute value being usually less than 0.5 in
the considered region), nucleons are assumed to obey Boltzmann 
statistics, and, normalized to representative values,
$M_{1.1}$ is measured in units of $1.1\,M_\odot$, $\rho_9$ in
$10^9\,{\rm g/cm}^3$, and $r_7$ in $10^7\,{\rm cm}$. Inserting
typical numbers ($T\approx 1.5\,{\rm MeV}$, $Y_n \approx 0.3$,
$R_{\rm cut}\approx 1.5\cdot 10^7\,{\rm cm}$), one obtains
$s > 15\,k_{\rm B}/N$, which gives
an estimate of the entropy in the heating region when the star
is going to explode.

These requirements can be coupled to the neutrino emission of the 
proto-neutron star by the following considerations.
A stalled shock is converted into a moving
one only when the neutrino heating is strong enough to increase the pressure
behind the shock by a sufficient amount. Considering the Rankine-Hugoniot
relations at the shock, Bruenn (1993) derived a criterion for the heating rate 
per unit mass, $q_{\nu}$, behind the shock that guarantees a positive postshock
velocity ($u_1 > 0$):
\begin{equation}
q_{\nu}\,>\,{2\beta - 1\over \beta^3(\beta-1)(\gamma-1)}\,
{|u_0|^3 \over \eta R_{\rm s}} \, .
\label{eq-5}
\end{equation}
Here $\beta$ is the ratio of postshock to preshock density,
$\beta = \rho_1/\rho_0$, $\gamma$ the adiabatic index of the gas
(assumed to be the same in front and behind the shock), and $\eta$
defines the fraction of the shock radius $R_{\rm s}$ where net heating
by neutrino processes occurs: $\eta = (R_{\rm s}-R_{\rm g})/R_{\rm s}$.
$u_0$ is the preshock velocity, which is a fraction $\alpha$
(analytical and numerical calculations show that typically
$\alpha \approx 1/\sqrt{2}$) of the free fall velocity,
$u_0 = \alpha\sqrt{2GM/r}$. Assuming a strong shock, one has 
$\beta = (\gamma+1)/(\gamma-1)$ which becomes $\beta = 7$ for 
$\gamma = 4/3$. With numbers typical of the collapsed core of the 
$15\,M_{\odot}$ star considered by Janka \& M\"uller (1996), 
$R_{\rm s} = 200\,{\rm km}$, $\eta\approx 0.4$, and an interior mass
$M = 1.1\,M_\odot$, one finds for the threshold luminosities of 
$\nu_e$ and $\bar\nu_e$:
\begin{equation}
L_{\nu,52}\langle\epsilon_{\nu,15}^2\rangle\ >\ 2.0\,\,
{M_{1.1}^{3/2}\over R_{{\rm s},200}^{1/2}} \, .
\label{eq-6}
\end{equation}
The existence of such a threshold luminosity of the order of 
$2\cdot 10^{52}\,$erg/s is underlined by Fig.~\ref{fig-2} where the
explosion energy $E_{>0}$ as function of time is shown for numerical 
calculations of the same post-collapse model but with different
assumed neutrino luminosities from the proto-neutron star. 
$E_{>0}$ is defined to include the sum of internal, kinetic,
and gravitational energy for all zones where this sum is positive 
(the gravitational binding energies of stellar mantle and envelope
and additional energy release from nuclear burning are not taken into
account). For 
one-dimensional simulations with luminosities below $1.9\cdot 10^{52}\,$erg/s
we could not get explosions when the proto-neutron star was assumed static,
and the threshold for the $\nu_e$ and $\bar\nu_e$ luminosities was
$2.2\cdot 10^{52}\,$erg/s when the neutron star was contracting
(see Janka \& M\"uller 1996). The supporting effects of convective overturn
between the gain radius and the shock described above lead to explosions 
even below the
threshold luminosities for the spherically symmetric case, to higher 
values of the explosion energy for the same neutrino luminosities, and
to a faster development of the explosion. This can clearly be seen by 
comparing the solid (2D) and dashed (1D) lines in Fig.~\ref{fig-2}.

The results of Fig.~\ref{fig-2} also show that the explosion energy
is extremely sensitive to the neutrino luminosities and mean energies. 
This holds in 1D as well as in 2D. Dick McCray in his summary talk of
this conference raised the question why neutrino-driven explosions 
should be self-regulated. Which kind of feedback should prevent the 
explosion from being more energetic than a few times $10^{51}\,$erg?
Certainly, the neutrino luminosities in current models can hardly power
an explosion and therefore a way to overpower it is not easy to imagine.
Nevertheless, Fig.~\ref{fig-2} and Eq.~(\ref{eq-2}) offer an answer to
Dick's question: When the matter in the neutrino-heated region outside
the gain radius has absorbed roughly its gravitational binding energy 
from the neutrino fluxes, it starts to expand outward (see Eq.~(\ref{eq-3}))
and moves away from the region of
strongest heating. Since the onset of the explosion shuts off the re-supply
of the heating region with cool gas, the curves in 
Fig.~\ref{fig-2} approach a saturation level as soon as the expansion
gains momentum and the density in the heating region decreases. Thus 
the explosion energy depends on the strength of the neutrino heating, which
scales with the $\nu_e$ and $\bar\nu_e$ luminosities and mean energies, and 
it is limited by the amount of matter $\Delta M$ in the heating
region and by the duration of the heating (see Eq.~(\ref{eq-2})), both 
of which decrease when the heating is strong and expansion happens
fast. 

This also implies that neutrino-driven explosions can be ``delayed''
(up to a few 100~ms after core bounce) but are {\it not} ``late''
(after a few seconds) explosions. The density between the gain radius
and the shock decreases with time because the proto-neutron star 
contracts and the mass infall onto the collapsed core declines steeply
with time. Therefore the mass $\Delta M$ in the heated region drops 
rapidly and energetic explosions by the neutrino-heating mechanism
become less favored at late times.

\begin{figure}
\plotone{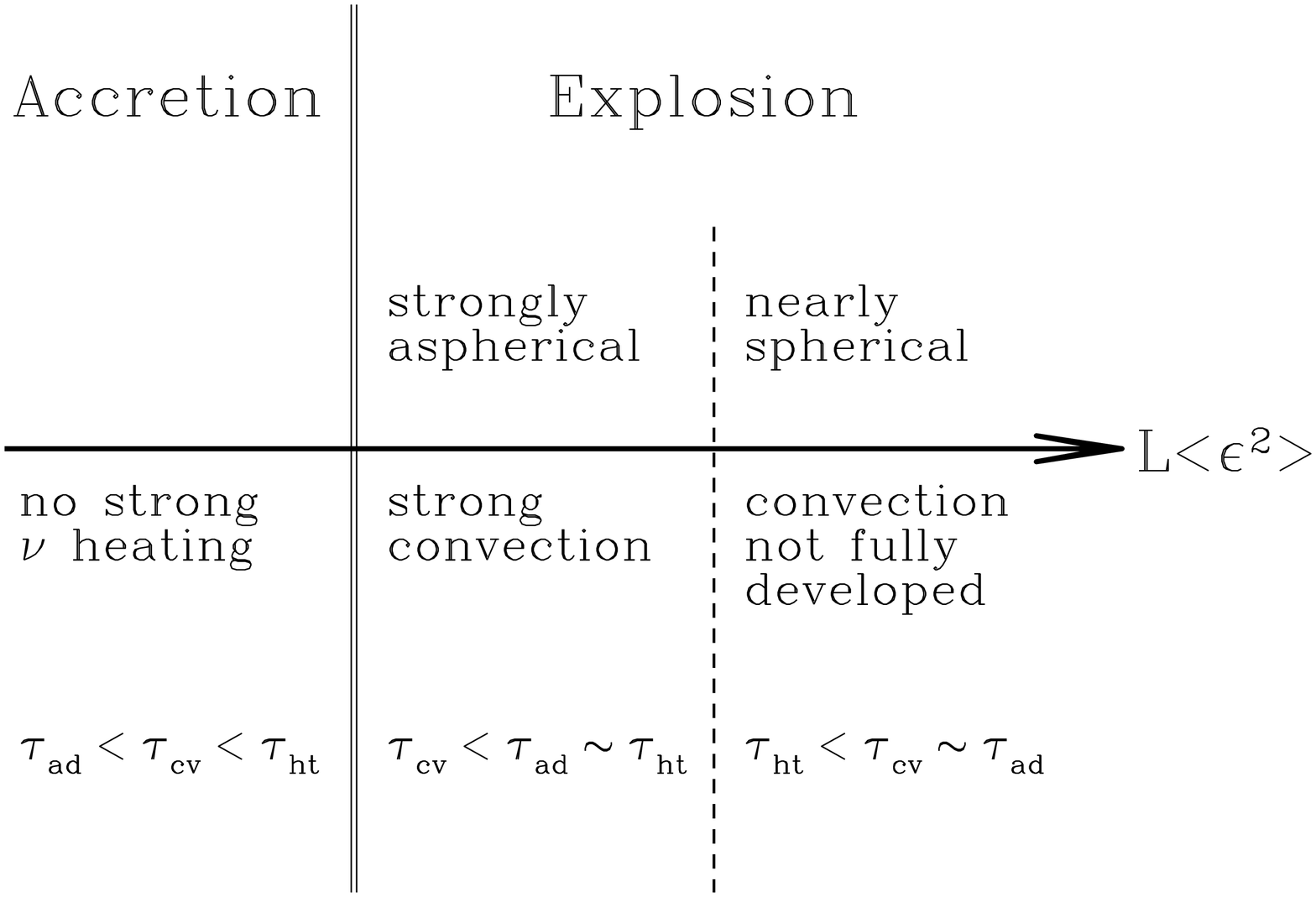}
\caption{Order scheme for the dependence of the post-collapse dynamics
on the strength of the neutrino heating as a function of 
$L_{\nu}\langle\epsilon_{\nu}^2\rangle$.}
\label{fig-3}
\end{figure}

Moreover, Fig.~\ref{fig-2} tells us that convection is {\it not necessary} 
to get an explosion and convective overturn is {\it no guarantee} for strong
explosions. Therefore one must suspect that neutrino-driven type-II 
explosions should reveal a considerable spread in the explosion energies,
even for similar progenitor stars. Rotation in the stellar core, small
differences of the core mass or statistical variations in the dynamical
events that precede and accompany the explosion may lead to some variability. 

The role of convective overturn and its importance for the explosion can
be further illuminated by considering the three timescales of neutrino heating,
$\tau_{\rm ht}$, advection of accreted matter through the gain radius 
into the cooling region and onto the neutron star (compare Fig.~\ref{fig-1}),
$\tau_{\rm ad}$, and growth of convective overturn, $\tau_{\rm cv}$.
The evolution of the shock --- accretion or explosion --- is determined by the
relative sizes of these three timescales. Straightforward considerations
show that they are of the same order and the destiny of the star is therefore 
a result of a tight competition between the different processes
(see Fig.~\ref{fig-3}). 

The heating timescale is estimated from the initial
entropy $s_{\rm i}$, the critical entropy $s_{\rm c}$ (Eq.~(\ref{eq-4})),
and the heating rate per nucleon (Eq.~(\ref{eq-1})) as
\begin{equation}
\tau_{\rm ht}
\,\approx\,{s_{\rm c}-s_{\rm i}\over Q_{\nu}^+/(k_{\rm B}T)}
\,\approx\,45\,{\rm ms}\cdot {s_{\rm c}-s_{\rm i}\over 5 k_{\rm B}/N}
\,{R_{{\rm g},7}^2(T/2{\rm MeV})\,f\over(L_{\nu}/2\cdot 10^{52}{\rm erg/s})
\langle\epsilon_{\nu,15}^2\rangle} \; .
\label{eq-7}
\end{equation}
With a postshock velocity of 
$u_1 = u_0/\beta \approx (\gamma-1)\sqrt{GM/R_{\rm s}}/(\gamma+1)$ 
the advection timescale is
\begin{equation}
\tau_{\rm ad}\,\approx\,
{R_{\rm s}-R_{\rm g}\over u_1}\,\approx\,52\,{\rm ms}\cdot 
\left(1-{R_{\rm g}\over R_{\rm s}}\right)\,{R_{{\rm s},200}^{3/2}\over
\sqrt{M_{1.1}}} \; ,
\label{eq-8}
\end{equation}
where the gain radius can be determined as
\begin{equation}
R_{{\rm g},7}\,\cong \,0.4 
\left( {L_{\nu}\over 2\cdot 10^{52}{\rm erg/s}}\right)^{-1/4}
\langle\epsilon_{\nu,15}^2\rangle^{-1/4} f^{1/4}
\left( {R_{\rm ns}\over 25 {\rm km}}\right)^{3/2}
\label{eq-9}
\end{equation}
from the requirement that the heating rate, Eq.~(\ref{eq-1}), is equal to 
the cooling rate per nucleon, $Q_{\nu}^-\approx 288(T/2{\rm MeV})^6$, when
use is made of the power-law behavior of the temperature according to 
$T(r)\approx 4\,{\rm MeV}\,(R_{\rm ns}/r)$ with $R_{\rm ns}$ being the 
proto-neutron star radius (roughly equal to the neutrinosphere radius).
The growth timescale 
of convective instabilities in the neutrino-heated region depends on the
gradients of entropy and lepton number through the growth rate of 
Ledoux convection, $\sigma_{\rm L}$ ($g$ is the gravitational acceleration):
\begin{equation}
\tau_{\rm cv}\,\approx\,{\ln{(100)}\over \sigma_{\rm L}}\,\approx\,
4.6\left\lbrace{g\over\rho}\left\lbrack
\left({\partial\rho\over\partial s}\right)_{\! Y_e,P}
{{\rm d}s\over{\rm d}r}+\left({\partial\rho\over\partial Y_e}\right)_{\! s,P}
{{\rm d}Y_e\over{\rm d}r}
\right\rbrack\right\rbrace^{-1/2}\!\gtrsim\,50\,{\rm ms} \; .
\label{eq-10}
\end{equation}
The numerical value is representative for those obtained in hydrodynamical 
simulations (e.g., Janka \& M\"uller 1996). $\tau_{\rm cv}$ of Eq.~(\ref{eq-10})
is sensitive to the detailed conditions between neutrinosphere (where 
$Y_e$ has typically a minimum), gain radius (where $s$ develops a maximum),
and the shock. The neutrino heating timescale is shorter for larger values
of the neutrino luminosity $L_{\nu}$ and mean squared neutrino energy 
$\langle\epsilon_{\nu}^2\rangle$, while both $\tau_{\rm ht}$ and $\tau_{\rm ad}$
depend strongly on the gain radius, $\tau_{\rm ad}$ also on the shock
position.

\begin{figure}
\plottwo{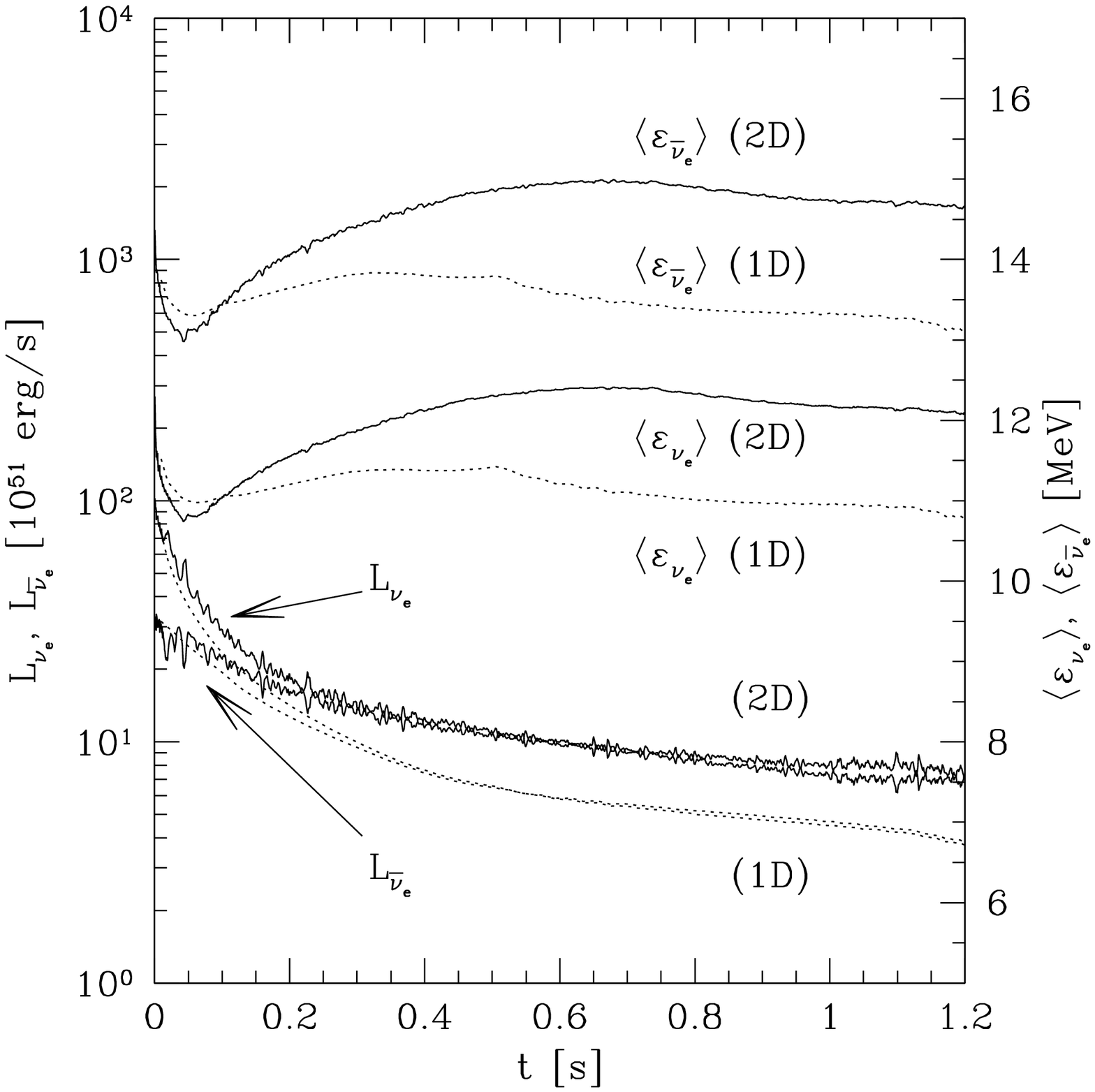}{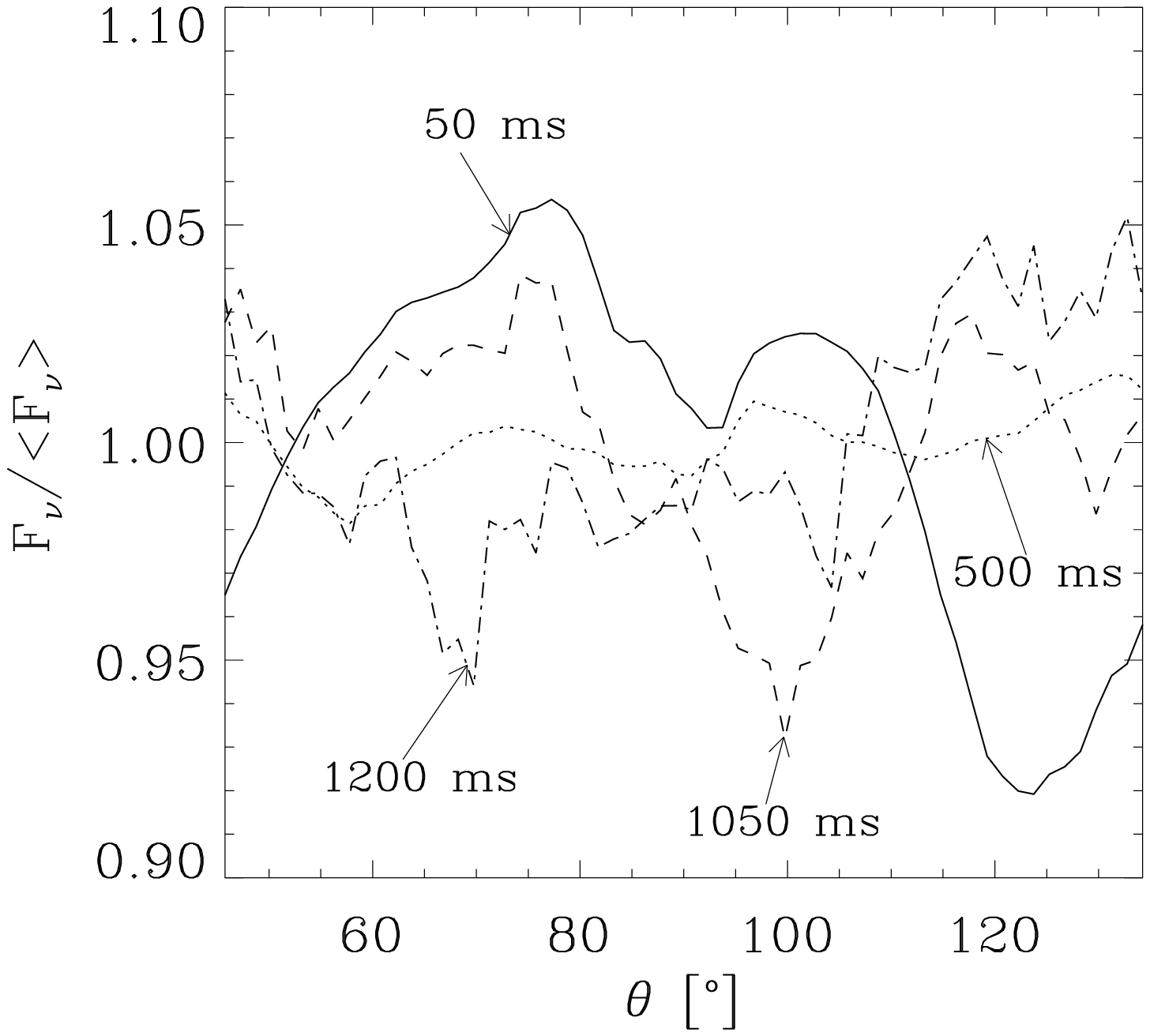}
\caption{Left: Luminosities $L_{\nu}(t)$ and mean energies 
$\langle\epsilon_{\nu}\rangle(t)$ of $\nu_e$ and $\bar\nu_e$ for a
$1.1\,M_{\odot}$ proto-neutron star without (``1D''; dotted) and with convection 
(``2D''; solid). Time is measured from core bounce.
Right: Angular variations of the neutrino flux at 
different times for the 2D simulation.}
\label{fig-4}
\end{figure}

\section{Convection inside the nascent neutron star}\label{sec-3}

Convective energy transport inside the newly formed neutron star can increase
the neutrino luminosities considerably (Burrows 1987). This could be
crucial for energizing the stalled supernova shock (Mayle \& Wilson 1988;
Wilson \& Mayle 1988, 1993; see also Sect.~\ref{sec-2}).

Recent two-dimensional simulations by Keil et al.~(1996, 1997) and Keil (1997)
have followed the evolution of the proto-neutron 
star formed in the core collapse of a $15\,M_{\odot}$ star for a period of 
more than 1.2~seconds. The simulations were performed with the hydrodynamics
code {\it Prometheus}. A general relativistic 1D gravitational potential 
with Newtonian corrections for asphericities was used, 
$\Phi\equiv\Phi_{\rm 1D}^{\rm GR}+(\Phi_{\rm 2D}^{\rm N}-\Phi_{\rm 1D}^{\rm N})$,
and a flux-limited (equilibrium) neutrino diffusion scheme was applied
for each angular bin separately (``1${1\over 2}$D'').

The simulations show that convectively unstable surface-near regions 
(i.e., around the neutrinosphere and below an initial density of about
$10^{12}\,$g/cm$^3$) exist only for a short period of a few ten 
milliseconds after bounce, in agreement with the findings of 
Bruenn \& Mezzacappa (1994), Bruenn et al.~(1995), and Mezzacappa et al.~(1997).
Due to a flat entropy profile and a negative lepton number gradient,
convection, however, also starts in a layer deeper inside the star, 
between an enclosed mass of $0.7\,M_{\odot}$ and $0.9\,M_{\odot}$, at
densities above several $10^{12}\,$g/cm$^3$. From there the convective 
region digs into the star and reaches the center after about one second.
Convective velocities as high as $5\cdot 10^8\,$cm/s are reached 
(about 10--20\% of the local sound speed), corresponding to kinetic 
energies of up to 1--$2\cdot 10^{50}\,$erg. Because of these high velocities 
and rather flat entropy and composition profiles in the star, the overshoot
(and undershoot) regions are large. 

The coherence lengths of convective structures are of the order of 20--40
degrees (in 2D!) and coherence times are of the order of 10~ms which
corresponds to only one or two overturns. The convective pattern is 
therefore very time-dependent and nonstationary.
Convective motions lead to considerable variations of the composition.
The lepton fraction (and thus the abundance of protons) shows relative 
fluctuations of several 10\%. The entropy differences in rising and sinking
convective bubbles are much smaller, only a few per cent, while temperature
and density fluctuations are typically less than one per cent. 

The energy transport in the neutron star is dominated by neutrino diffusion
near the center, whereas convective transport plays the major role in a thick 
intermediate layer where the convective activity is strongest, and radiative
transport takes over again when the neutrino mean free path becomes large
near the surface of the star. But even in the convective layer the convective
energy flux is only a few times larger than the diffusive flux. This means
that neutrino diffusion can never be neglected. 

There is an important consequence of this latter statement. The convective
activity in the neutron star cannot be described and explained as Ledoux
convection. Applying the Ledoux criterion for local instability,
$C_{\rm L}(r) = (\rho/g)\sigma_{\rm L}^2 > 0$ with $\sigma_{\rm L}$ from
Eq.~(\ref{eq-10}) and $Y_e$ replaced by the total lepton fraction 
$Y_{\rm lep}$ in the neutrino-opaque interior of the neutron star, 
one finds that the convecting region should actually be stable, despite
of slightly negative entropy {\it and} lepton number gradients. In fact, 
below a critical value of the lepton fraction (e.g., 
$Y_{\rm lep,c} = 0.148$ for $\rho = 10^{13}\,$g/cm$^3$ and $T = 10.7\,$MeV)
the thermodynamical derivative $(\partial \rho/\partial Y_{\rm lep})_{s,P}$
changes sign and becomes positive because of nuclear and Coulomb forces in
the high-density equation of state. Therefore negative lepton number gradients
should stabilize against convection in this regime. However, an idealized
assumption of Ledoux convection is not fulfilled in the situations considered
here: Because of neutrino diffusion energy exchange and, in particular,
lepton number exchange between convective elements and their surroundings 
are {\it not} negligible. Taking the neutrino transport effects on 
$Y_{\rm lep}$ into account in a modified  {\it Quasi-Ledoux criterion} 
(Keil 1997 and Keil et al.~1997) one predicts instability 
exactly where convective action happens in the two-dimensional simulation.

\section{Conclusions}\label{sec-4}

Convection inside the proto-neutron star can raise the neutrino luminosities 
within a few hundred ms after core bounce (Fig.~\ref{fig-4}).
In the considered collapsed core of a $15\,M_{\odot}$ star $L_{\nu_e}$ and
$L_{\bar\nu_e}$ increase by up to 50\% and the mean neutrino energies by 
about 15\% at times later than 200--300~ms post bounce.
This favors neutrino-driven explosions on timescales of a few
hundred milliseconds after shock formation. Also, the deleptonization of
the nascent neutron star is strongly accelerated, raising the $\nu_e$ 
luminosities relative to the $\bar\nu_e$ luminosities during this time. 
This helps to increase the electron fraction $Y_e$ in the neutrino-heated
ejecta and might solve the overproduction problem of $N=50$ nuclei 
during the early epochs of the explosion
(Keil et al.~1996). Anisotropic mass motions due to convection in the 
neutron star lead to gravitational wave emission and anisotropic radiation
of neutrinos. The angular variations of the neutrino flux determined by the 
2D simulations are of the order of 5--10\% (Fig.~\ref{fig-4}). With the typical
size and short coherence times of the convective structures, however, the 
global anisotropy of the neutrino emission from the cooling proto-neutron
star is certainly less than 1\% (more likely only 0.1\%, since in 3D the
structures tend to be smaller) and kick velocities in excess of 300~km/s 
can definitely not be explained. 

In more recent simulations, mass accretion and rotation of the forming
neutron star were included. Rotation has very interesting consequences,
e.g., leads to a suppression of convective motions near the rotation 
axis because of a stabilizing stratification of the specific angular
momentum, an effect which can be understood by applying the 
Solberg-H\o iland criterion for instabilities in rotating, self-gravitating
objects. Future simulations will have to clarify the influence of the
nuclear equation of state on the presence of convection in nascent 
neutron stars. Also, a more accurate treatment of the neutrino transport 
in combination with a state-of-the-art description of the neutrino 
opacities of the nuclear medium is needed to confirm the existence of
a convective episode during neutron star formation and to
study its importance for the explosion mechanism of type-II supernovae.

\acknowledgments

Many discussions and a fruitful collaboration with W.~Keil and E.~M\"uller
are acknowledged. The author is very grateful to the organizers for the
opportunity to participate in the Ten-Years-After meeting in La Serena
and is thankful that SN1987A has happened during a decisive phase of his
scientific life.

\end{document}